\documentclass[10pt,journal]{IEEEtran}

\usepackage{amsmath}
\usepackage{amsfonts}
\usepackage{subcaption}
\usepackage{graphicx}
\usepackage{float}
\usepackage{tabularx}
\usepackage{amssymb}
\usepackage{pdflscape}  
\usepackage{setspace}
\usepackage{color}          
\usepackage{graphics} 
\usepackage{subfig}                      
\usepackage{epsfig} 
\usepackage{times} 
\usepackage[hyphens]{url}  
\usepackage{comment}
\usepackage{algorithmicx,algorithm}
\usepackage{cite}
\usepackage{booktabs}
\usepackage{makecell}
\usepackage{multirow}

\title{Headroom-Aware Stochastic Adaptive Model Predictive Control for Load Frequency Control in Microgrids}

\author{
Erfan Mehdipour Abadi,
Shuo Yuan,
Le Yi Wang,
Caisheng Wang,
and Feng Lin%
\thanks{Erfan Mehdipour Abadi, Shuo Yuan, Le Yi Wang, Caisheng Wang, and Feng Lin are with the Department of Electrical and Computer Engineering, Wayne State University, Detroit, MI 48202 USA. }%
\thanks{Emails: { \{erfan.abadi, shuoyuan, lywang, cwang, aa0986\}}@wayne.edu.}
\thanks{Corresponding author: Le Yi Wang.}%
}

\date{ }

\begin{document}

\maketitle
\begin{abstract}

As the penetration of inverter-based resources (IBRs) increases in microgrids, they are increasingly expected to play a greater role in load frequency control (LFC). Model predictive control (MPC) is attractive for LFC because it incorporates system dynamics and operational constraints. However, most MPC-based LFC formulations rely on fixed reserve headroom based on forecasted renewable availability or storage systems. Under short-term renewable intermittency, IBR headroom is stochastic and time-varying, causing optimal control commands to exceed the physically deliverable regulation capability and cause stochastic saturation. This control-actuator mismatch degrades LFC performance. Accordingly, this study develops two headroom-aware strategies for PV-dominated microgrids. First, stochastic headroom constrained MPC (SHCMPC) incorporates headroom predictions through time-varying input constraints to enforce control feasibility. Second, stochastic adaptive MPC (SAMPC) embeds headroom awareness into the MPC objective function by adaptively penalizing control actions based on predicted headroom, reducing reliance on units with limited headroom without hard time-varying constraints. Simulation results show that stochastic saturation degrades conventional MPC-based LFC, particularly under tight reserve margins. Both strategies improve regulation performance. SHCMPC eliminates saturation events, while SAMPC achieves substantial saturation mitigation with lower computational effort, offering a computationally efficient alternative for real-time LFC under stochastic renewable availability.

\end{abstract}


\section{Introduction}

Microgrids are increasingly deployed to improve the flexibility, resilience, and sustainability of modern power systems \cite{guo2025co}. At the same time, high penetration of renewable resources such as photovoltaic and wind
generation is making inverter-based resources (IBRs) dominant in many emerging microgrids, motivating new control and monitoring schemes
\cite{abadi2025detecting, abadi2024distributed}. Since IBRs are interfaced through power electronic converters, they can provide fast and flexible power control, making them attractive not only as generation resources but also as providers of fast regulation services \cite{mo2025multi, cetinkaya2022impact}.

However, high IBR penetration also creates challenges for frequency regulation, which is the focus of this work.
The reduction of system inertia makes microgrid frequency dynamics more
sensitive to active-power imbalances, increasing the importance of load
frequency control (LFC)
\cite{mo2025multi, cetinkaya2022impact,
yusuf2024review, he2024analysis}. Since relying primarily on synchronous
generators or battery energy storage systems for regulation may be costly or
impractical in IBR-dominated microgrids
\cite{magableh2025tool, oshnoei2020robust}, IBRs
themselves can serve as LFC providers \cite{al2023overview}. Nevertheless, to provide upward reserve, IBRs must be scheduled below their forecasted available generation, creating headroom for frequency support at the expense of renewable curtailment; however, such reserve would otherwise need to be supplied by other resources with their own associated costs and limitations \cite{batzelis2018pv}.
The difference between forecasted capacity and scheduled output defines the
scheduled reserve margin. Unlike conventional reserve sources, however,
renewable IBR headroom is inherently uncertain and stochastically time-varying
because it depends on environmental conditions
\cite{samal2025load, yusuf2024review}. Therefore, this paper investigates how
stochastic renewable IBR headroom affects the feasibility of MPC-based LFC
commands and develops headroom-aware MPC strategies to mitigate the resulting
stochastic saturation.

Many existing LFC formulations are built around predictable or controllable
regulation resources, whose capacities are modeled by fixed or forecasted
limits. Similarly, conventional fixed-headroom LFC designs assume that the
scheduled reserve margin of IBRs remains available during real-time operation
\cite{al2023overview}. Although reducing this margin improves renewable
utilization, it also makes LFC more vulnerable to short-term availability
uncertainty. If the actual headroom falls below scheduled margin, the LFC
command may exceed the instantaneous deliverable power and be clipped by the
physical availability limit. This optimization-actuation mismatch is referred to
in this work as \emph{stochastic saturation}, whose impact becomes more
pronounced under reduced reserve margins.

A wide range of LFC strategies has been developed for modern power systems
and microgrids, including classical, optimal, robust, fuzzy, sliding-mode, and
learning-based approaches \cite{yusuf2024review, gulzar2025load,
abbasi2025lfc}. In systems with high IBR penetration, reduced inertia and the
displacement of conventional resources increase the need for controllers that
can exploit fast active-power regulation from IBRs
\cite{al2023overview, he2024analysis}. Among these approaches, model predictive
control (MPC) has attracted particular attention because it optimizes control
actions over a prediction horizon while explicitly incorporating system
dynamics, operational constraints, and performance objectives
\cite{yi2020distributed, joshal2023microgrids, hu2021model}. MPC-based
frequency control has been studied for systems with renewable generation,
storage devices, and distributed energy resources (DERs), where renewable uncertainty
is often modeled as an uncertain power injection \cite{wang2024frequency, taher2023optimal, MPC_Primary}, or renewable/IBR
participation is represented through deterministic input constraints,
forecast-based limits, or probabilistic reserve ranges
\cite{ademola2020frequency, ma2024distributed, yi2020distributed,
zhang2025secondary}. However, these formulations do not explicitly analyze how
reducing the scheduled reserve margin affects MPC-based LFC through stochastic
headroom saturation. Thus, the short-term mismatch between scheduled reserve
margin and actual available headroom remains insufficiently addressed.

Adaptive and learning-based MPC formulations have also been proposed for LFC
under renewable uncertainty and time-varying operating conditions
\cite{zeng2017adaptive, yang2020inertia, zhao2022adaptive, negahban2022novel,
wang2024frequency, zhang2023adaptive ,abadi2026contingency}. These methods typically adapt the
prediction model, controller parameters, weights, horizons, or constraints to
improve robustness against model and operating-condition variations. However,
they do not specifically address the short-term stochastic mismatch between the
scheduled reserve margin and the actual renewable headroom. In our prior work
\cite{StochasticAdaptiveDroop}, IBR generation capacity was modeled as a
Markovian stochastic process and used to adapt droop coefficients based on
real-time availability. Extending this idea to MPC-based LFC enables explicit
consideration of headroom-dependent available power, reserve-margin reduction,
and stochastic saturation within a predictive optimal control framework.

In this paper, stochastic headroom-aware MPC frameworks for LFC in IBR-dominated microgrids are introduced. The proposed frameworks explicitly analyze how short-term renewable availability affects the feasibility of LFC commands and show that stochastic saturation can significantly degrade frequency regulation, especially as the scheduled reserve margin decreases. By exploiting the short-term predictability of stochastic renewable availability modeled as Markovian processes, the proposed controllers incorporate predicted headroom information into MPC-based LFC and improve regulation performance under tight reserve conditions.

The main contributions of this paper are summarized as follows:
\begin{itemize}
    \item A stochastic headroom-based formulation of the LFC problem is developed for the two proposed MPC strategies. The formulation distinguishes between scheduled reserve margin and actual available headroom, characterizes the resulting stochastic saturation, and quantifies its impact on frequency regulation performance using standard LFC indices and saturation-severity measures.

    \item A stochastic headroom constrained MPC (SHCMPC) strategy is proposed, in which predicted short-term renewable headrooms are incorporated as time-varying input constraints. By explicitly enforcing headroom-dependent feasibility, SHCMPC mitigates the optimization-actuation mismatch caused by stochastic saturation and improves LFC performance under reduced reserve margins.

    \item A stochastic adaptive MPC strategy is developed as an alternative headroom-aware control approach. Instead of imposing predicted headrooms as explicit time-varying constraints, SAMPC incorporates normalized headroom information through adaptive input penalties, redistributing control effort away from resources with limited available headrooms while preserving the standard MPC structure. Comparative case studies show that SAMPC achieves effective frequency regulation under reduced reserve margins with lower real-time computational burden.
\end{itemize}

The remainder of this paper is organized as follows: Section~\ref{Sec_2}
formulates the problem; Section~\ref{Sec_3} develops the microgrid and
stochastic headroom models; Section~\ref{Sec_4} presents the conventional
MPC, SHCMPC, and SAMPC formulations; Section~\ref{Sec_6} provides case
studies; and Section~\ref{Sec_7} concludes the paper.

\section{Problem Formulation} \label{Sec_2}

The objective of LFC is to regulate the microgrid frequency $f(t)$ around its rated value $f_{\mathrm{ref}}$ by correcting power imbalance between generation and demand. The frequency deviation is defined as
$\Delta f(t) = f(t) - f_{\mathrm{ref}}$ which is to be regulated. In this work, frequency regulation is provided by $N$ renewable IBRs, whose power outputs are denoted by $P_{G,i}(t)$, $i=1,\ldots, N$, without ESSs or conventional generators. Hence, the regulation reserve is determined by the headrooms of the renewable IBRs. The total load demand of the microgrid $P_L(t)$ is modeled as an exogenous disturbance to the frequency dynamics. 
In this study, uncertain load variations are modeled as disturbance inputs. In contrast, renewable generation uncertainty is modeled through the IBR headrooms and directly modifies the feasible ranges of the control inputs. This modeling distinction motivates the focus on stochastic headroom uncertainty.

At the dispatching layer, the energy management system computes a nominal operating point from renewable generation and load demand forecasts. Let $\hat{P}_i^{\mathrm{f}}$ denote the forecasted generation capacity of the $i$th IBR, and let $P_i^{\mathrm{nom}}$ denote its scheduled nominal power. To enable upward regulation, the IBR is scheduled below its forecasted capacity, i.e., $P_i^{\mathrm{nom}} < \hat{P}_i^{\mathrm{f}}$, creating the scheduled reserve margin
\begin{equation*}
    R_i^{\mathrm{sch}} = \hat{P}_i^{\mathrm{f}} - P_i^{\mathrm{nom}}, \quad i=1, \dots, N.
\end{equation*}
This quantity represents the forecast-based headroom assumed available for LFC in fixed-headroom conventional MPC designs. Increasing the dispatch fraction $P_i^{\mathrm{nom}}/\hat{P}_i^{\mathrm{f}}$ improves renewable utilization but reduces $R_i^{\mathrm{sch}}$, making the reserve margin a control-relevant dispatch parameter that directly affects LFC capability.

The scheduled outputs define the equilibrium point around which the linear state-space LFC model is derived. Accordingly, for each interval, the dispatching layer is assumed to select $P_i^{\mathrm{nom}}$ such that $\sum_{i=1}^{N} P_i^{\mathrm{nom}} = P_L^{\mathrm{nom}},$
where $P_L^{\mathrm{nom}}$ is the total power demand, including load demand and power losses, associated with that operating point. Therefore, changing the reserve margin corresponds to selecting a different feasible equilibrium point, and the LFC controller regulates deviations around that equilibrium.

In practice, IBR available power is not fixed at its forecasted value. Let $P_i^{\max}(t)$ denote the instantaneous maximum power that can be delivered by the IBR. Due to IBR intermittency, $P_i^{\max}(t)$ is stochastic and time varying, and deviates from the forecasted $\hat{P}_i^{\mathrm{f}}$. The actual available headroom for LFC is
\begin{equation} \label{ActualHeadroom}
    H_i(t) = P_i^{\max}(t) - P_i^{\mathrm{nom}}.
\end{equation}
Unlike the scheduled reserve margin $R_i^{\mathrm{sch}}$, the headroom $H_i(t)$ represents the real-time upward regulation capability of the IBR. In general, $H_i(t) \neq R_i^{\mathrm{sch}}$, and the mismatch becomes more critical as $H_i(t)$ decreases. The case $H_i(t)<0$ indicates that the IBR cannot fully deliver its nominal power due to severe short-term availability loss, let alone provide upward regulation reserve.

Given a nominal equilibrium point, LFC acts on real-time deviations from the scheduled IBR outputs. The LFC control input of the $i$th IBR is defined as the active-power deviation
\begin{equation*}
    u_i(t) \triangleq P_{G,i}(t)-P_i^{\mathrm{nom}},
\end{equation*}
where $P_{G,i}(t)$ is the real-time power injection. Since the IBR output cannot exceed its instantaneous available power or become negative, the physically admissible range of $u_i(t)$ is
\begin{equation}
    -P_i^{\mathrm{nom}} \leq u_i(t) \leq H_i(t),
\end{equation}
where the upper bound is determined by the stochastic headroom, $H_i(t)$, while the lower bound corresponds to zero power injection, $P_{G,i}(t)=0$. Thus, the feasible control set of each IBR is time-varying and stochastic, even though the nominal operating point is fixed during the dispatch interval. For analytical clarity, this paper is focused on headroom-induced input feasibility and additional device-level constraints such as inverter ramp-rate limits are not explicitly modeled.

Conventional MPC-based LFC designs commonly use offline-designed input bounds based on the scheduled reserve margin. In this case, the admissible upward regulation is represented as
\begin{equation}
    -P_i^{\mathrm{nom}} \leq u_i(t) \leq R_i^{\mathrm{sch}},
\end{equation}
rather than the real-time feasible bound $H_i(t)$. Therefore, an MPC command may satisfy $u_i(t)\leq R_i^{\mathrm{sch}}$ while violating the physical constraint $u_i(t)\leq H_i(t)$ when $H_i(t)<R_i^{\mathrm{sch}}$.
To distinguish the controller-requested regulation command from the
regulation power that can actually be delivered under stochastic
headroom limitations, we denote the MPC-computed command by
$u_i^{\mathrm{cmd}}(t)$ and the physically applied input by
$u_i^{\mathrm{act}}(t)$.
During implementation, the applied input is clipped by the instantaneous availability limit, i.e.,
\begin{equation}\label{eq:Uactual}
    u_i^{\mathrm{act}}(t)=\min\{u_i^{\mathrm{cmd}}(t),H_i(t)\},
\end{equation}
with the lower bound enforced similarly. This clipping creates an optimization-actuation mismatch between the MPC command and the delivered IBR power, namely \emph{stochastic saturation}. This mismatch becomes more influential under reduced reserve margins because the limited headroom makes the condition $H_i(t)<R_i^{\mathrm{sch}}$ more likely to result in input clipping and LFC performance degradation.

Based on this formulation, next section develops two modeling components required for headroom-aware MPC-based LFC. First, an aggregate linear frequency response model is presented from the active-power balance equation around the nominal equilibrium point to predict the frequency response over the MPC horizon. Second, a Markovian stochastic model is introduced to characterize and predict short-term variations of renewable IBR headrooms. These components are then used to construct the SHCMPC and SAMPC formulations.

\section{Stochastic Headroom-Aware Microgrid Modeling} \label{Sec_3}
This section develops the modeling framework used for stochastic headroom-aware MPC in an IBR-dominated microgrid. First, the microgrid control architecture is described. Then, an aggregate frequency-response model is derived from active-power balance and written in state-space form for MPC prediction. Finally, IBR availability is modeled as a Markovian stochastic process to characterize and predict short-term headroom variations.

\subsection{Microgrid Control Architecture}
\begin{figure}[t]
    \includegraphics[width=1\linewidth, trim={1cm 1cm 1cm 1cm},clip]{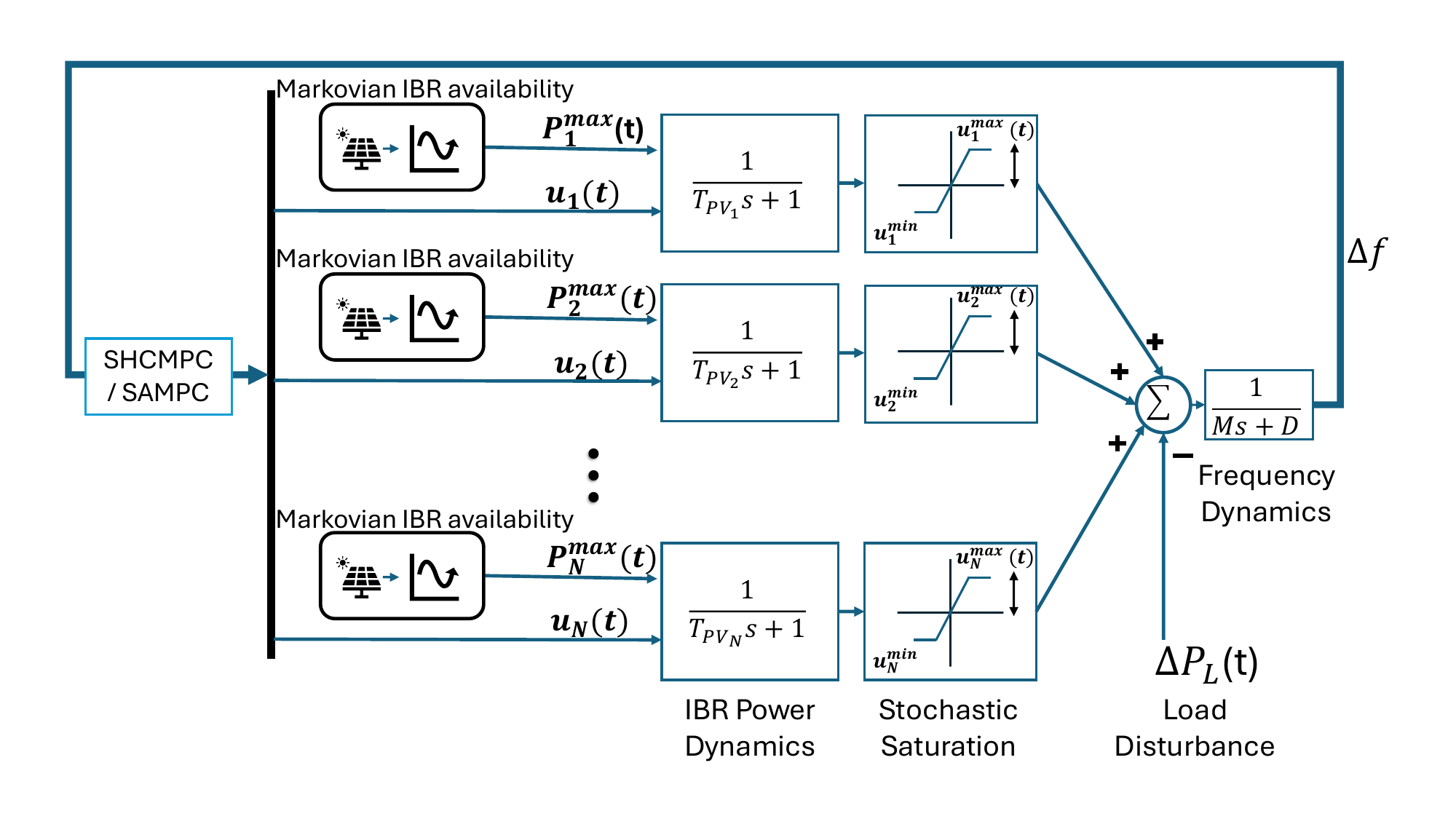}
        \centering
        \captionsetup{width=\linewidth}
        \caption{Structure of the IBR-dominated microgrid with stochastic headroom constraints and an MPC-based LFC layer.}
        \label{Fig_SystemStructure}
        \vspace{-1.5em}
\end{figure}

Fig.~\ref{Fig_SystemStructure} illustrates the control architecture of the IBR-dominated microgrid considered in this study. The microgrid includes $N$ renewable IBR units, represented here by PV systems interfaced through power-electronic converters. The formulation, however, applies to other IBRs whose available powers determine their real-time regulation headrooms.

Each IBR receives an LFC command $u_i(t)$, defined as the active-power deviation from its nominal scheduled output. The converter-level active-power response is modeled by first-order dynamics with time constant $T_{\mathrm{pv},i}$. The instantaneous availability $P_i^{\max}(t)$ is modeled as a stochastic process and determines the headroom $H_i(t)$. Accordingly, the stochastic saturation block limits the delivered command using $u_i^{\max}(t)=H_i(t)$ and $u_i^{\min}=-P_i^{\mathrm{nom}}$.

The saturated active-power deviations are aggregated and applied to the microgrid frequency-response dynamics. The load deviation $\Delta P_L(t)$ enters as an exogenous disturbance, while the resulting frequency deviation $\Delta f(t)$ is fed back to the headroom-aware LFC layer. The SHCMPC and SAMPC controllers use $\Delta f(t)$ and short-term headroom information to compute the IBR regulation commands.

\subsection{Frequency Response Model}

To formulate MPC-based LFC, an aggregate frequency-response model is adopted, in which the frequency deviation is driven by the net power imbalance\cite{wang2024frequency}. The dynamics are:
\begin{equation}
    \dot{\Delta f} = -\frac{D}{M}\Delta f
    + \frac{1}{M}\sum_{i=1}^{N} z_{\mathrm{pv},i}
    - \frac{1}{M}\Delta P_L,
    \label{eq_PowerBalance}
\end{equation}
where $M>0$ and $D\geq 0$ denote the equivalent inertia and damping coefficients, respectively. $\Delta P_L(t)=P_L(t)-P_L^{\mathrm{nom}}$ is the load deviation, and $z_{\mathrm{pv},i}$ denotes the realized power deviation of the $i$th PV unit.

Each PV unit is represented by first-order power dynamics that capture converter and power-regulation delays:
\begin{equation}
    \dot{z}_{\mathrm{pv},i}
    =
    -\frac{1}{T_{\mathrm{pv},i}}z_{\mathrm{pv},i}
    +\frac{1}{T_{\mathrm{pv},i}}u_i,
    \label{eq_pvDynamic}
\end{equation}
where $T_{\mathrm{pv},i}>0$ is the time constant and $u_i$ is the power deviation command generated by LFC for the $i$th PV unit.

Define the continuous-time state vector, control input, and load disturbance as
\begin{align*}
    x_c &\triangleq 
    \begin{bmatrix}
        \Delta f & z_{\mathrm{pv},1} & \cdots & z_{\mathrm{pv},N}
    \end{bmatrix}^{\!\top},\\
    u &\triangleq 
    \begin{bmatrix}
        u_1 & \cdots & u_N
    \end{bmatrix}^{\!\top}, 
    \qquad
    d \triangleq \Delta P_L,
\end{align*}
with measured output \(y\triangleq \Delta f\). The resulting continuous-time state-space model is
\begin{subequations}\label{eq:cts_ss}
\begin{align}
    \dot{x}_c &= A_c x_c + B_c u + E_c d,\\
    y &= C_c x_c,
\end{align}
\end{subequations}
where
\begin{align*}
A_c &=
\begin{bmatrix}
-\tfrac{D}{M} & \tfrac{1}{M}\mathbf{1}_N^{\top} \\
\mathbf{0}_{N\times 1} & -\Theta
\end{bmatrix}, 
&
B_c &=
\begin{bmatrix}
\mathbf{0}_{1\times N} \\
\Theta
\end{bmatrix}, \nonumber\\
E_c &=
\begin{bmatrix}
-\tfrac{1}{M}\\
\mathbf{0}_{N\times 1}
\end{bmatrix}, 
&
C_c &=
\begin{bmatrix}
1 & \mathbf{0}_{1\times N}
\end{bmatrix}.
\label{eq:cts_matrices}
\end{align*}
Here, \(\mathbf{1}_N\) is the \(N\)-dimensional all-ones column vector, \(\mathbf{0}_{m\times n}\) is the \(m\times n\) zero matrix, and
\begin{equation*}
\Theta \triangleq \mathrm{diag}\!\left(\frac{1}{T_{\mathrm{pv},1}},\ldots,\frac{1}{T_{\mathrm{pv},N}}\right).
\end{equation*}

Since the MPC is implemented at discrete samples, \eqref{eq:cts_ss} is discretized by using zero-order hold (ZOH) with sampling period $T_s>0$ as
\begin{align}
x_c(k+1) &= A_d x_c(k) + B_d u(k) + E_d d(k),\\
y(k) &= C_d x_c(k),
\end{align}
where $A_d=e^{A_cT_s}$, $B_d=(\int_0^{T_s}e^{A_cs}ds)B_c$, $E_d=(\int_0^{T_s}e^{A_cs}ds)E_c$, and $C_d=C_c$.

The discrete-time model captures the frequency response to PV power deviations and load disturbances. While penalizing $y(k)=\Delta f(k)$ improves frequency regulation, persistent disturbances may lead to steady-state frequency error without integral action. To incorporate secondary-frequency-regulation behavior, an integral state of the frequency deviation is added as
\begin{equation}
    e(k+1)=e(k)+T_s y(k).
    \label{eq:integrator_state}
\end{equation}
Define the augmented state as
\begin{equation*}
    x(k) \triangleq \begin{bmatrix}x_c^\top(k) & e(k)\end{bmatrix}^{\top}.
\end{equation*}
The augmented discrete-time model is
\begin{subequations}\label{eq:aug_ss}
\begin{align}
    x(k+1) &= A x(k)+B u(k)+E d(k),\\
    y(k) &= C x(k),
\end{align}
\end{subequations}
where
\begin{align*}
A&=\begin{bmatrix}A_d & \mathbf{0}\\ T_s C_d & 1\end{bmatrix}, \quad
B=\begin{bmatrix}B_d\\ \mathbf{0}_{1\times N}\end{bmatrix}, \nonumber\\
E&=\begin{bmatrix}E_d\\0\end{bmatrix}, \quad
C=\begin{bmatrix}C_d & 0\end{bmatrix}.
\end{align*}
The model in \eqref{eq:aug_ss} will be used as the MPC prediction model in the subsequent control formulations.

\subsection{Markovian Stochastic Model for IBR Capacity}
The real-time IBR capacity $P_i^{\max}(k)$ varies with environmental conditions such as irradiance, cloud coverage, temperature, and shading. Although these variations are stochastic, they are temporally correlated over short prediction horizons; therefore, the capacity does not change independently from one sampling instant to the next. Following the stochastic availability modeling approach in \cite{StochasticAdaptiveDroop}, $P_i^{\max}(k)$ is quantized into discrete capacity levels and modeled as an \emph{infrequently switching} finite-state Markov chain.

An infrequently switching Markov chain is a discrete-time stochastic process whose transition probabilities are dominated by self-transitions, so the process tends to remain in its current state with high probability over short horizons. The transition probability matrix of the $i$th IBR capacity process is expressed as
\begin{equation} \label{MarkovianTransitionMatrix}
    P_i = I + \varepsilon G_i,
\end{equation}
where $I$ is the identity matrix of appropriate dimension, $0<\varepsilon\ll 1$ characterizes the rate of capacity variation, and $G_i$ is a generator matrix satisfying
\[
[G_i]_{ab}\ge 0\ (a\ne b),\qquad \sum_b [G_i]_{ab}=0,\ \forall a
\]
where the model parameters can be estimated from historical or measured available-power data using standard statistical methods. 
In this study, the Markov chain is used to provide short-term prediction $P_i^{\max}(t)$ over MPC horizon. Thus, the realizations of this capacity process determine the stochastic headroom defined in \eqref{ActualHeadroom}, which is used to construct the headroom-aware MPC formulations. However, the proposed headroom-aware MPC framework is not restricted to Markovian prediction, and any alternative forecasting method providing $\hat{P}_i^{max}(k+h|k)$, can be used for headroom-aware MPC.

After the IBR capacity is realized at time $k$, the deliverable upward regulation is limited by the actual headroom $H_i(k)$. Hence, any LFC command exceeding this limit is clipped according to the applied input relation in \eqref{eq:Uactual}. The actual plant evolution is therefore described by \eqref{eq:aug_ss} with the commanded input $u(k)$ replaced by the applied input $u^{\mathrm{act}}(k)$. The next section uses the frequency-response model and stochastic headroom representation to develop the headroom-aware MPC-based LFC frameworks.

\section{Headroom-Aware MPC-Based LFC}\label{Sec_4}

This section develops the MPC-based LFC formulations using the system dynamics model in \eqref{eq:aug_ss}. At each sampling instant, a finite-horizon optimization problem is solved, and only the first control action is applied in a receding-horizon manner. The control objective is designed to regulate frequency deviation and remove steady-state frequency error by penalizing the integral state, while the input constraints characterize the available regulation capability of the PV units. Although additional constraints can be incorporated within the MPC framework, this work focuses on headroom-induced input feasibility. The distinction among the formulations lies in how this feasibility is represented: conventional MPC uses offline-designed fixed bounds based on the scheduled reserve margin, SHCMPC uses predicted stochastic headroom as time-varying constraints, and SAMPC introduces real-time stochastic adaptation through headroom-dependent input penalties.

\subsection{Conventional Constrained MPC-Based LFC}

In this work, the conventional MPC baseline refers to a constrained MPC formulation in which the input limits are fixed offline according to the scheduled reserve margin. Therefore, although the baseline MPC includes input constraints, these constraints are not updated according to the stochastic real-time headroom availability.

Conventional MPC-based LFC is considered as the fixed-headroom baseline. At each sampling instant, the controller uses the state-space model in \eqref{eq:aug_ss} to predict frequency evolution over a prediction horizon \(N_p\), computes PV active-power commands over a control horizon \(N_c\), and applies the first command in a receding-horizon manner. The optimization minimizes frequency deviation, accumulated frequency error, and excessive input modulation, while enforcing offline-designed fixed input bounds based on the scheduled reserve margin \(R_i^{\mathrm{sch}}\). This formulation provides the reference case against which the proposed headroom-aware MPC strategies are evaluated.

In the following MPC formulations, the notation $(k+h|k)$ denotes
the predicted value at future time $k+h$ computed at time $k$ using
the information available at the current sampling instant.
Accordingly, $x(k+h|k)$ denotes the predicted state, and
$u(k+h|k)$ denotes the future commanded control input over the
prediction horizon.
The finite-horizon MPC cost is defined as
\begin{align} \label{objectiveFunction}
    J &= \sum_{h=1}^{N_p} x(k+h|k)^\top Q\,x(k+h|k) \notag \\
      &\quad + \sum_{h=0}^{N_c-1} u(k+h|k)^\top R\,u(k+h|k),
\end{align}
where $N_p$ and $N_c\leq N_p$ denote the prediction and control horizons, respectively. The first term penalizes the predicted frequency deviations and accumulated frequency errors, while the second term regularizes PV power modulation.

The state weighting matrix $Q\succeq0$ is selected to penalize $\Delta f(k)$ and the integral state $e(k)$, while leaving the internal PV power-response states unpenalized. Since the augmented state is ordered as $x=[\Delta f,z_{\mathrm{pv},1},\ldots,z_{\mathrm{pv},N},e]^\top$, $Q$ is chosen as
\begin{equation}
    Q \triangleq \mathrm{diag}\!\left(q_f,\mathbf{0}_{1\times N},q_I\right),
    \label{eq:Qdef}
\end{equation}
where $q_f>0$ and $q_I>0$ are the weights on frequency deviation and accumulated frequency error, respectively. The relative values of $q_f$ and $q_I$ tune the balance between transient frequency regulation and steady-state error removal; in this work, $q_f>q_I$ is used to avoid excessive emphasis on the integral state which could lead to unwanted oscillations.

The input weighting matrix is selected as
\begin{equation}
    R \triangleq \lambda I_N,
    \label{eq:Rdef}
\end{equation}
where $\lambda>0$ imposes a uniform penalty on PV control inputs. This quadratic objective is consistent with standard MPC-based LFC formulations \cite{wang2024frequency}.

For compact implementation of the finite-horizon MPC problem, the system dynamics in \eqref{eq:aug_ss} are written in a lifted form. The predicted state trajectory is expressed as a function of the current state and the future control sequence. Define
\[
X \triangleq 
\begin{bmatrix} 
x(k+1|k) \\ x(k+2|k) \\ \vdots \\ x(k+N_p|k) 
\end{bmatrix}, 
\qquad
U \triangleq 
\begin{bmatrix} 
u(k|k) \\ u(k+1|k) \\ \vdots \\ u(k+N_c-1|k) 
\end{bmatrix}.
\]
Then,
\begin{equation}
    X = \Phi x(k) + \Gamma U,
    \label{eq:lifted}
\end{equation}
where
\[
\Phi =
\begin{bmatrix}
A \\ A^2 \\ \vdots \\ A^{N_p}
\end{bmatrix},
\qquad
\Gamma =
\begin{bmatrix}
B & 0 & \cdots & 0 \\
A B & B & \cdots & 0 \\
\vdots & \vdots & \ddots & \vdots \\
A^{N_p-1}B & A^{N_p-2}B & \cdots & B
\end{bmatrix}.
\]
The disturbance sequence is not included in the lifted prediction model since future load deviations are unknown. Their effect is handled through receding-horizon feedback, where the measured state is updated at each sampling instant.

Using the lifted prediction model in \eqref{eq:lifted}, the finite-horizon cost in \eqref{objectiveFunction} can be expressed compactly in terms of the stacked input vector \(U\). For this purpose, define the block-diagonal weighting matrices
\[
\mathcal{Q} \triangleq I_{N_p}\otimes Q,
\qquad
\mathcal{R} \triangleq I_{N_c}\otimes R.
\]
where $\otimes$ denotes the Kronecker product, $I_{N_p}$ and $I_{N_c}$ are identity matrices of appropriate dimensions. Similarly, the fixed input bounds of the conventional MPC formulation are stacked over the control horizon as
\[
U_{\min} \triangleq \mathbf{1}_{N_c}\otimes u^{\min},
\qquad
U_{\max}^{\mathrm{sch}} \triangleq \mathbf{1}_{N_c}\otimes u_{\max}^{\mathrm{sch}},
\]
where $\mathbf{1}_{N_c}$ is an $N_c$-dimensional column vector of ones and
\[
u^{\min} \triangleq
\begin{bmatrix}
-P_1^{\mathrm{nom}} \\ \vdots \\ -P_N^{\mathrm{nom}}
\end{bmatrix},
\qquad
u_{\max}^{\mathrm{sch}} \triangleq
\begin{bmatrix}
R_1^{\mathrm{sch}} \\ \vdots \\ R_N^{\mathrm{sch}}
\end{bmatrix}.
\]

The conventional constrained MPC-based LFC problem at time $k$ is:
\begin{subequations}\label{ConventionalMPC}
\begin{align}
\min_{U}\quad 
& J(k)=X^\top \mathcal{Q}X+U^\top \mathcal{R}U \label{ConventionalMPC1}\\
\text{s.t.}\quad
& X=\Phi x(k)+\Gamma U, \label{ConventionalMPC2}\\
& U_{\min}\leq U\leq U_{\max}^{\mathrm{sch}}, \label{ConventionalMPC3}
\end{align}
\end{subequations}
with constraints $U_{\min}$ and $ U_{\max}^{\mathrm{sch}}$.

After solving \eqref{ConventionalMPC}, the first element of the optimal sequence is applied as the LFC command in a receding-horizon manner. Since the upper bound $U_{\max}^{\mathrm{sch}}$ is fixed from the scheduled reserve margin, this baseline does not update input feasibility according to the actual available headroom. Therefore, when the actual plant is driven by the applied input relation in \eqref{eq:Uactual}, commands that exceed the actual headrooms are clipped, which can degrade LFC performance under reduced reserve margins. The next subsection modifies the MPC constraints to explicitly incorporate predicted stochastic headroom.

\subsection{Stochastic Headroom Constrained MPC}
SHCMPC modifies the conventional MPC baseline by replacing offline-designed fixed input bounds with predicted stochastic headroom bounds. Following the Markovian capacity model in Section~\ref{Sec_3}, short-term PV capacity predictions are converted into predicted headroom values and used to construct time-varying upper input bounds over the control horizon. Thus, unlike conventional MPC, which constrains LFC commands by the scheduled reserve margin, SHCMPC constrains them by the predicted  headroom at each future step.

Let $\hat{P}_i^{\max}(k+h|k)$ denote the predicted available capacity of the $i$th PV unit at time $k+h$ based on information available at time $k$, for $h=0,\ldots,N_c-1$. The corresponding predicted headroom is
\begin{equation*}
    \hat{H}_i(k+h|k)
    =
    \hat{P}_i^{\max}(k+h|k)-P_i^{\mathrm{nom}} .
    \label{eq:predicted_headroom}
\end{equation*}
To account for the higher reliability of short-term headroom predictions, stochastic headroom constraints are enforced over the first \(N_c^{\mathrm{hard}}\leq N_c\) control moves. For the remaining moves, the scheduled reserve bound is used. Thus, the SHCMPC upper-bound vector at step \(k+h\) is defined as
\begin{equation}
u_{\max}^{\mathrm{SHC}}(k+h|k)=
\begin{cases}
\hat{H}(k+h|k), & h=0,\ldots,N_c^{\mathrm{hard}}-1,\\
u_{\max}^{\mathrm{sch}}, & h=N_c^{\mathrm{hard}},\ldots,N_c-1,
\end{cases}
\label{eq:sh_mixed_bound}
\end{equation}
where \(\hat{H}(k+h|k)=[\hat{H}_1(k+h|k),\ldots,\hat{H}_N(k+h|k)]^\top\) and \(u_{\max}^{\mathrm{sch}}=[R_1^{\mathrm{sch}},\ldots,R_N^{\mathrm{sch}}]^\top\).

Stacking these bounds over the control horizon gives
\begin{equation}
U_{\max}^{\mathrm{SHC}}(k) \triangleq
\begin{bmatrix}
u_{\max}^{\mathrm{SHC}}(k|k)\\
u_{\max}^{\mathrm{SHC}}(k+1|k)\\
\vdots\\
u_{\max}^{\mathrm{SHC}}(k+N_c-1|k)
\end{bmatrix}.
\label{eq:Umax_pred}
\end{equation}

Hence, the SHCMPC problem at time $k$ is formulated as
\begin{subequations}\label{SHMPC}
\begin{align}
\min_{U}\quad & J(k)=X^\top \mathcal{Q}X+U^\top \mathcal{R}U \\
\text{s.t.}\quad
& X=\Phi x(k)+\Gamma U,\\
& U_{\min}\leq U\leq U_{\max}^{\mathrm{SHC}}(k).
\end{align}
\end{subequations}
Compared with \eqref{ConventionalMPC}, the fixed upper bound $U_{\max}^{\mathrm{sch}}$ is replaced by the predicted stochastic headroom bound $U_{\max}^{\mathrm{SHC}}(k)$. This constrains the optimized LFC commands by predicted deliverable PV headroom rather than by the scheduled reserve margin, reducing the likelihood of stochastic saturation. However, SHCMPC enforces headroom awareness through explicit time-varying input constraints, requiring the solution of a computationally more demanding constrained QP at each sampling instant. The next subsection develops SAMPC as an alternative formulation that embeds stochastic headroom information into adaptive input penalties, yielding an unconstrained MPC problem with lower computation time and making it appropriate for real-time LFC.

\subsection{Stochastic Adaptive MPC}

SAMPC is developed as an unconstrained headroom-aware MPC formulation in which stochastic PV availability is reflected through adaptive input weighting rather than time-varying hard constraints. In a standard MPC formulation with uniform input penalties, control effort is allocated primarily according to the system dynamics and input effectiveness. In SAMPC, the input penalty is updated according to the predicted headroom of each PV unit, reducing reliance on units with limited available headroom while preserving an unconstrained quadratic optimization structure.

For each prediction step \(k+h\), \(h=0,\ldots,N_c-1\), the predicted headroom vector \(\hat{H}(k+h|k)\) is used to construct an adaptive input-weighting matrix. The predicted headroom values are normalized to quantify the relative share of available regulation capability associated with each PV unit. Units with smaller normalized headroom shares are assigned larger penalties, while units with larger shares are assigned smaller penalties, thereby shifting regulation effort toward PV units with greater available headroom. The adaptation also includes an aggregate-headroom scarcity factor, which increases the overall input penalty when the total predicted regulation capability becomes limited. Negative predicted headroom values indicate that the available capacity is below the nominal scheduled output and therefore result in stronger penalization of capacity-deficient units. Let 

\begin{equation*}
    \hat{H}_{\Sigma}(k+h|k)
    =
    \sum_{i=1}^{N}\hat{H}_i(k+h|k).
\end{equation*}
Thus, the normalized headroom share of the $i$th PV unit is defined as
\begin{equation*}
    \rho_i(k+h|k)
    =
    \frac{\hat{H}_i(k+h|k)}
    {\hat{H}_{\Sigma}(k+h|k)+\epsilon_H},
\end{equation*}
where $\epsilon_H>0$ is a small regularization constant.

First, define the relative adaptive factor
\begin{equation*}
    \tilde r_i(k+h|k)
    =
    \exp\!\left(-\kappa \rho_i(k+h|k)\right),
    \label{eq:relative_penalty}
\end{equation*}
where \(\kappa>0\) controls the sensitivity to relative headroom. To preserve the average input-penalty level while redistributing effort among PV units, the relative factors are normalized as
\begin{equation*}
    \bar r_i(k+h|k)
    =
    \frac{\tilde r_i(k+h|k)}
    {\frac{1}{N}\sum_{j=1}^{N}\tilde r_j(k+h|k)} .
    \label{eq:normalized_relative_penalty}
\end{equation*}
The final adaptive penalty is selected as
\begin{equation}
    r_i(k+h|k)
    =
    \lambda_{\mathrm{base}}\,
    \gamma(k+h|k)\,
    \bar r_i(k+h|k),
    \label{eq:adaptive_penalty}
\end{equation}
where \(\lambda_{\mathrm{base}}>0\) is the baseline input penalty and \(\gamma(k+h|k)\geq 1\) is an aggregate-headroom scarcity factor.

The scarcity factor is defined as
\begin{equation}
\gamma(k+h|k)
=
\exp\!\left(
\kappa_g
\max\left\{
0,\,
\frac{\bar H^{\mathrm{sch}}-\bar H(k+h|k)}
{\bar H^{\mathrm{sch}}+\epsilon_H}
\right\}
\right),
\label{eq:scarcity_factor}
\end{equation}
where
\[
\bar H(k+h|k)=\frac{1}{N}\sum_{i=1}^{N}\hat H_i(k+h|k)
\]
is the predicted average headroom and \(\bar H^{\mathrm{sch}}\) is the scheduled average headroom. The normalized factor \(\bar r_i(k+h|k)\) redistributes regulation effort according to relative headroom availability, while \(\gamma(k+h|k)\) increases the overall penalty when aggregate predicted headroom becomes scarce. Thus, SAMPC remains close to nominal MPC behavior under sufficient reserve and becomes more conservative when available regulation capability is limited.

In contrast to the fixed uniform input-weighting matrix $R$ used in the conventional MPC formulation in \eqref{eq:Rdef}, SAMPC replaces it with a headroom-dependent adaptive input-weighting matrix. Collecting the scalar adaptive penalties in \eqref{eq:adaptive_penalty}, the input-weighting matrix at prediction step $k+h$ is defined as
\begin{equation*}
R_a(k+h|k)
\triangleq
\mathrm{diag}\!\left(
r_1(k+h|k),\ldots,r_N(k+h|k)
\right),
\label{eq:Ra_step}
\end{equation*}
for $h=0,\ldots,N_c-1$. Stacking these matrices gives
\begin{equation*}
\mathcal{R}_a(k)\triangleq
\begin{bmatrix}
R_a(k|k) & \textbf{0} & \cdots &  \textbf{0}\\
 \textbf{0} & R_a(k+1|k) & \cdots & \textbf{0}\\
\vdots & \vdots & \ddots & \vdots\\
 \textbf{0} &  \textbf{0} & \cdots & R_a(k+N_c-1|k)
\end{bmatrix}
\label{eq:Ra_block}
\end{equation*}
The resulting block-diagonal matrix $R_a(k)$ replaces the fixed input-weighting matrix used in conventional MPC. Hence, the SAMPC optimization problem at time $k$ is formulated as
\begin{subequations}\label{SAMPC}
\begin{align}
\min_{U}\quad 
& J_a(k)=X^\top \mathcal{Q}X+U^\top \mathcal{R}_a(k)U \\
\text{s.t.}\quad
& X=\Phi x(k)+\Gamma U .
\end{align}
\end{subequations}
Compared with SHCMPC, SAMPC removes the explicit time-varying input constraints and incorporates stochastic headroom information through $R_a(k)$. Therefore, the optimization remains an unconstrained quadratic problem, which reduces online solution time while still discouraging control actions from PV units with limited predicted headroom. If the resulting command exceeds the realized headroom, the physical saturation relation in \eqref{eq:Uactual} is still applied at the plant level.

\section{Case Studies}
\label{Sec_6}
This section evaluates the proposed headroom-aware MPC formulations under stochastic PV availability. The simulations first examine the effect of stochastic saturation on conventional constrained MPC-based LFC, and then compare conventional MPC, SHCMPC, and SAMPC under different reserve margins. Since existing adaptive MPC approaches aim to address model or condition uncertainties rather than input feasibility, conventional MPC is used as the baseline to isolate the effect of stochastic headroom awareness. The evaluation focuses on LFC performance, saturation severity, and online computation time.

\subsection{Simulation Setup}

The simulation study considers the PV-dominated microgrid shown in Fig.~\ref{Fig_SystemStructure}, with \(N=4\) PV units participating in LFC. Each PV unit has a rated capacity of \(P_i^{\mathrm{rated}}=8~\mathrm{MW}\), and the planning-layer forecasted available capacity is assumed to be \(\hat{P}_i^{\mathrm{f}}=8~\mathrm{MW}\) for all \(i=1,\ldots,4\). The microgrid dynamics follow the aggregate frequency-response model in \eqref{eq:aug_ss}, with effective inertia \(M=0.20\) and damping coefficient \(D=0.012\). The PV converter dynamics are modeled by first-order lags with time constants \(T_{\mathrm{pv},i}=[1.53,\,1.71,\,1.89,\,2.07]~\mathrm{s}\), and the continuous-time model is discretized with sampling period \(T_s=0.1~\mathrm{s}\) for MPC implementation.

The real-time available PV capacity \(P_i^{\max}(k)\) is generated for each unit using a bounded random-walk realization of the Markovian capacity model in Section~\ref{Sec_3}. Each trajectory is initialized at the forecasted value, \(P_i^{\max}(0)=\hat{P}_i^{\mathrm{f}}\), and evolves within \(0\leq P_i^{\max}(k)\leq P_i^{\mathrm{rated}}\) using a quantization step of \(\Delta P=0.05~\mathrm{MW}\). At each sampling instant, the capacity remains unchanged with probability \(p_{\mathrm{stay}}=0.99\), while upward and downward transitions occur with probabilities \(p_{\mathrm{up}}=p_{\mathrm{down}}=0.005\). The nominal scheduled power is selected as \(P_i^{\mathrm{nom}}=\alpha \hat{P}_i^{\mathrm{f}}\), where \(\alpha\in\{0.80,0.90,0.95\}\), corresponding to scheduled reserve margins of \(20\%\), \(10\%\), and \(5\%\), respectively. The resulting stochastic headroom is \(H_i(k)=P_i^{\max}(k)-P_i^{\mathrm{nom}}\).

Since this study focuses on stochastic headroom rather than load uncertainty, the load variation is modeled as a deterministic step disturbance. A \(0.2~\mathrm{MW}\) load increase is applied at \(t=40~\mathrm{s}\), while PV capacities evolve stochastically according to the random-walk model. This allows the effect of headroom uncertainty on LFC performance to be evaluated under identical load conditions. For each reserve margin scenario, the dispatch layer is assumed to schedule the nominal PV outputs and nominal load demand consistently, such that $P_L^{\mathrm{nom}}=\sum_{i=1}^{N}P_i^{\mathrm{nom}}.$ Thus, each case starts from a balanced equilibrium point before the load disturbance is applied.

The MPC controllers are implemented with prediction horizon \(N_p=50\) and control horizon \(N_c=5\). For SHCMPC, stochastic headroom constraints are enforced over the first \(N_c^{\mathrm{hard}}=2\) control moves, while the remaining moves use the scheduled reserve bounds. The MPC weighting parameters are selected as \(q_f=10^5\) and \(q_I=10^2\). The baseline input penalty is set to \(\lambda=10^{-2}\) for all controllers. In SAMPC, this value serves as the baseline coefficient in the adaptive weighting law, with headroom-sensitivity parameter \(\kappa=7\) and aggregate-scarcity scaling as defined in \eqref{eq:scarcity_factor}.

The simulations include four cases: ideal conventional MPC, conventional MPC with stochastic saturation, SHCMPC, and SAMPC. The conventional MPC case uses fixed scheduled reserve bounds and does not apply physical stochastic saturation, while the stochastic-saturation case applies the same MPC commands through the actual headroom limits. This comparison isolates the effect of optimization-actuation mismatch under stochastic PV availability. The SHCMPC and SAMPC cases are then evaluated to assess the effectiveness of the proposed headroom-aware control strategies. The controllers are evaluated using frequency-regulation indices, saturation metrics, and online computational burden.

LFC performance is evaluated over the simulation
horizon $T$ using
\begin{align}
\begin{aligned}
\mathrm{IAE} &= \int_0^T |\Delta f(t)|\,dt, &
\mathrm{ISE} &= \int_0^T \Delta f^2(t)\,dt,\\
\mathrm{ITAE} &= \int_0^T t|\Delta f(t)|\,dt, &
\mathrm{ITSE} &= \int_0^T t\Delta f^2(t)\,dt .
\end{aligned}
\label{eq:LFC_metrics}
\end{align}
These indices quantify the accumulated frequency error, large excursions, and persistence of frequency deviations.
To quantify stochastic saturation occurrence, the saturation severity is defined as
\begin{equation}
S_{\mathrm{sev}}
=
\sum_{k=0}^{N_T}\sum_{i=1}^{N}
\max\!\left\{0,\,
u_i^{\mathrm{cmd}}(k)-H_i(k)
\right\},
\label{eq:sat_severity}
\end{equation}
where \(u_i^{\mathrm{cmd}}(k)\) is the command computed by the controller and \(H_i(k)\) is the realized stochastic headroom. The saturation duration is measured by
\begin{equation}
S_{\mathrm{dur}}
= T_s
\sum_{k=0}^{N_T}
\mathbb{I}\!\left(
\sum_{i=1}^{N}
\max\!\left\{0,\,
u_i^{\mathrm{cmd}}(k)-H_i(k)
\right\}
>0
\right),
\label{eq:sat_duration}
\end{equation}
which measures the total duration over which at least one PV
unit experiences upward saturation. Here, $\mathbb{I}(\cdot)$ denotes
the indicator function, which is equal to one when its argument
is true and zero otherwise. Finally, computational burden is quantified by the average online solution time per control step. These settings are used consistently across all reserve margin scenarios and controller comparisons.

\subsection{Effect of Stochastic Saturation on Conventional MPC}

\begin{figure}[t]
    \centering
    \includegraphics[width=\columnwidth]{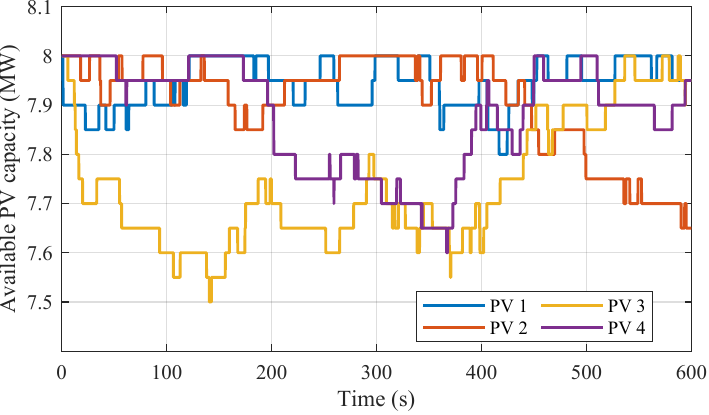}
    \caption{Illustrative stochastic available capacity realizations \(P_i^{\max}(k)\) of the four PV units generated using the Markovian model.}
    \label{fig:pv_avail}
    \vspace{-0.6cm}
\end{figure}

\begin{figure}[t]
    \centering
    \includegraphics[width=\columnwidth]{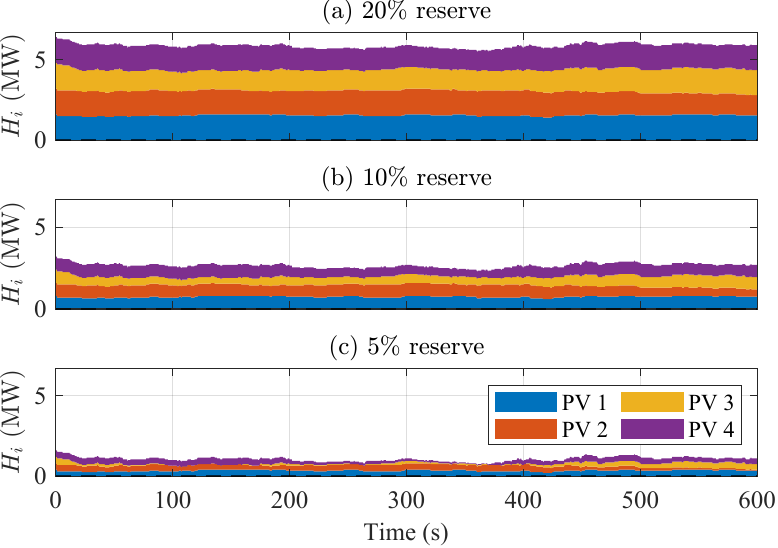}
    \caption{Stacked stochastic headroom profiles under (a) \(20\%\), (b) \(10\%\), and (c) \(5\%\) scheduled reserve margins.}
    \label{fig:headroom_cases}
\end{figure}

Before evaluating the control responses, the stochastic PV availability and the resulting headroom profiles are examined. Fig.~\ref{fig:pv_avail} illustrates the real-time available capacities \(P_i^{\max}(k)\) of the four PV units. The same realization of \(P_i^{\max}(k)\) is used for all reserve margin scenarios and control strategies to ensure a consistent comparison. Fig.~\ref{fig:headroom_cases} shows the corresponding headroom \(H_i(k)\) for the \(20\%\), \(10\%\), and \(5\%\) reserve-margin cases. Since the stochastic capacity realization is fixed, the headroom changes only through the nominal dispatch \(P_i^{\mathrm{nom}}=\alpha\hat P_i^{\mathrm{f}}\), with the nominal load adjusted accordingly to preserve the initial power balance. The stacked headroom profiles show both the aggregate regulation capability of the microgrid and the contribution of each PV unit.

As observed in Fig.~\ref{fig:headroom_cases}, stochastic headroom is unevenly
distributed among PV units, especially under the \(5\%\) reserve margin in
Fig.~\ref{fig:headroom_cases}(c). This shows that fixed scheduled reserve
margins do not capture the instantaneous distribution of available regulation
capability, motivating headroom-aware control allocation.

To quantify the effect of neglecting this variation, conventional MPC is
evaluated in two forms: an ideal implementation, where the optimized commands
are applied directly under fixed scheduled reserve bounds, and a stochastic
saturation implementation, where the same commands are clipped by the realized
headroom limits according to \eqref{eq:Uactual}. The difference between the two
responses represents the optimization-actuation mismatch caused by short-term
PV capacity variations.

\begin{figure}[t]
    \centering
    \includegraphics[width=\columnwidth]{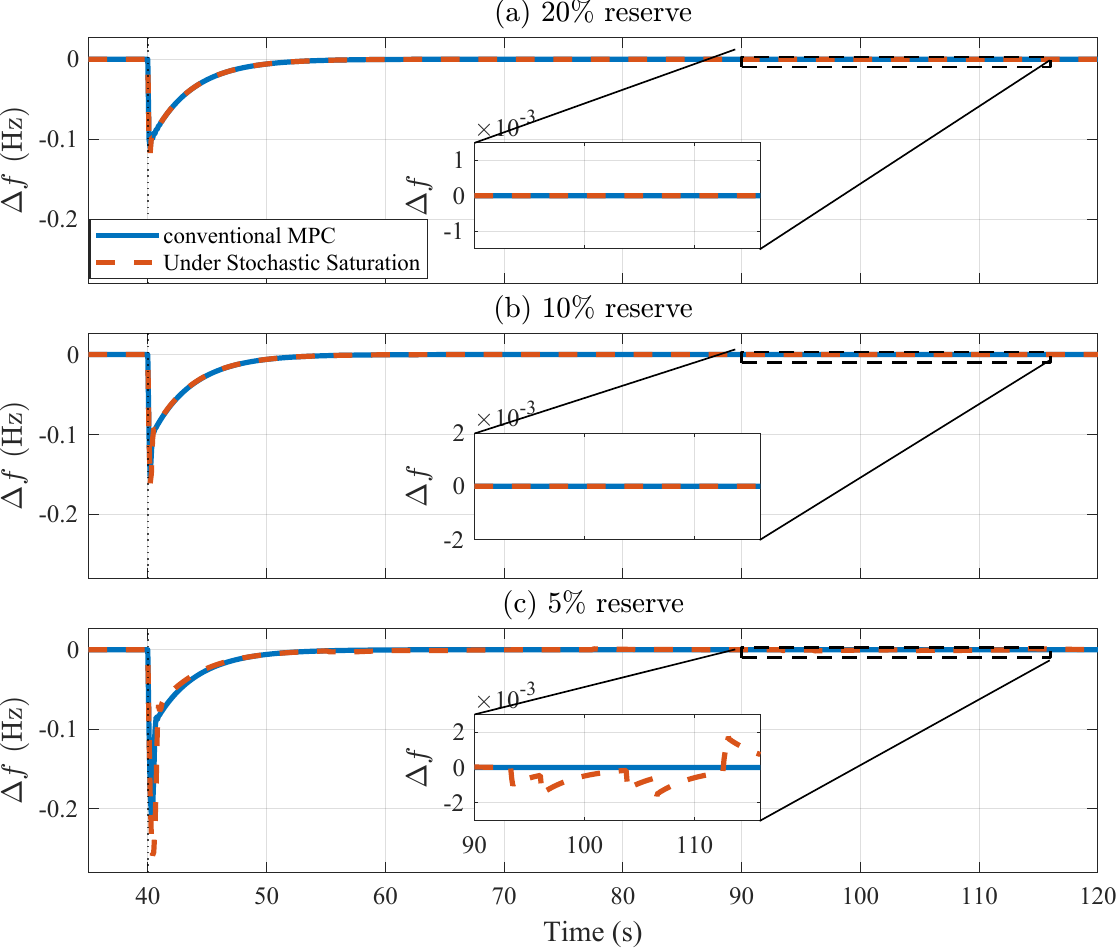}
    \caption{Frequency response of conventional MPC and conventional MPC under stochastic saturation within different reserve margins.}
    \label{fig:conv_sat_response}
    \vspace{-0.5cm}
\end{figure}

Fig.~\ref{fig:conv_sat_response} shows that conventional MPC restores the
frequency after the load disturbance in all reserve-margin cases. However, as
the scheduled reserve margin decreases, the frequency nadir becomes deeper and,
in the \(5\%\) reserve case, post-transient deviations remain visible. This
indicates that stochastic saturation mainly affects the sustained regulation
performance after the main recovery.

\begin{figure}[t]
    \centering
    \includegraphics[width=\columnwidth]{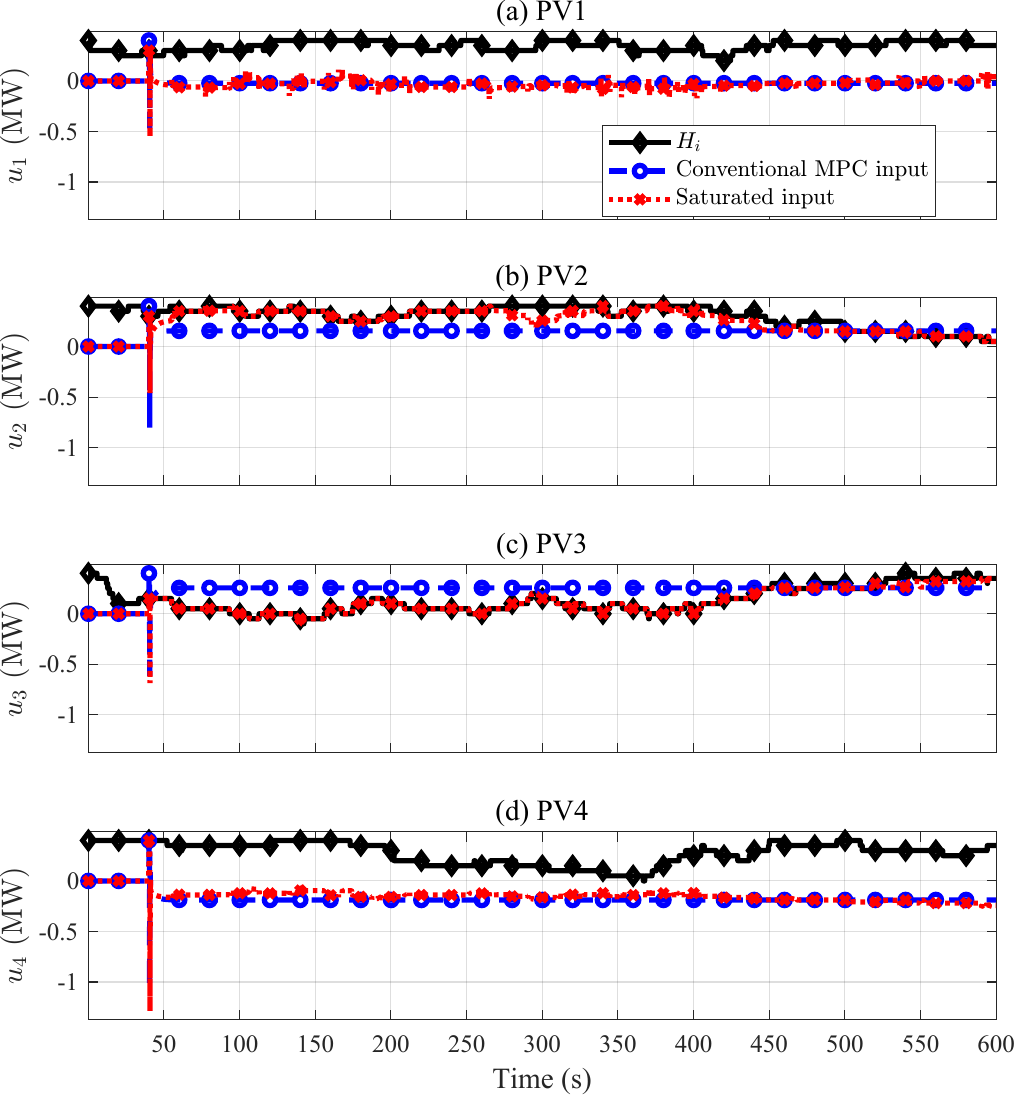}
    \caption{PV input saturation mechanism under the \(5\%\) reserve-margin scenario. The conventional MPC input is compared with the physically saturated input and the realized stochastic headroom bound.}
    \label{fig:input_saturation_5pct}
    \vspace{-0.4cm}
\end{figure}

Since this effect is most severe under limited reserve availability,
Fig.~\ref{fig:input_saturation_5pct} examines the input-level behavior for the
\(5\%\) reserve case. The conventional MPC command repeatedly exceeds the
realized headroom, especially for PV3 and later PV2, causing the delivered input
to be clipped by the saturation limit. Because this redistribution is triggered
indirectly through frequency feedback rather than planned using predicted
headroom, it occurs with delay and contributes to the post-transient
fluctuations in Fig.~\ref{fig:conv_sat_response}.

Table~\ref{tab:saturation_impact} confirms that stochastic saturation worsens
LFC performance, with degradation increasing as the reserve margin decreases.
The effect is strongest in the \(5\%\) reserve case, where stochastic
saturation increases IAE by approximately \(74\%\) and ITAE by approximately
\(594\%\). The larger ITAE increase indicates that the degradation is mainly
associated with persistent frequency deviations, consistent with
Fig.~\ref{fig:conv_sat_response}.

\begin{table}[t]
\centering
\caption{Impact of stochastic saturation on conventional MPC-based LFC.}
\label{tab:saturation_impact}
\setlength{\tabcolsep}{3.5pt}
\begin{tabular}{c l c c c c}
\toprule
\textbf{Reserve} & \textbf{Case} & \textbf{IAE} & \textbf{ISE} & \textbf{ITAE} & \textbf{ITSE} \\
\midrule
\multirow{2}{*}{20\%}
& Conventional & 0.3711 & 0.0200 & 16.14  & 0.8351 \\
& Stoch. Sat.  & 0.3711 & 0.0201 & 16.14  & 0.8384 \\
\midrule
\multirow{2}{*}{10\%}
& Conventional & 0.3849 & 0.0224 & 16.72  & 0.9331 \\
& Stoch. Sat.  & 0.3882 & 0.0231 & 17.19  & 0.9591 \\
\midrule
\multirow{2}{*}{5\%}
& Conventional & 0.4071 & 0.0305 & 17.60  & 1.2588 \\
& Stoch. Sat.  & 0.7100 & 0.0425 & 122.19 & 1.9009 \\
\bottomrule
\end{tabular}
\vspace{-0.7cm}
\end{table}

These results demonstrate that neglecting short-term headroom variations can lead to an optimistic assessment of conventional MPC-based LFC performance. In particular, under tight reserve margins, the mismatch between scheduled and realized regulation capability produces repeated input saturation and persistent frequency deviations. Therefore, the following subsection evaluates headroom-aware MPC strategies designed to incorporate stochastic PV availability into the control decision process.

\subsection{Performance of Headroom-Aware LFC Strategies}

\begin{figure}[t]
    \centering
    \includegraphics[width=\columnwidth]{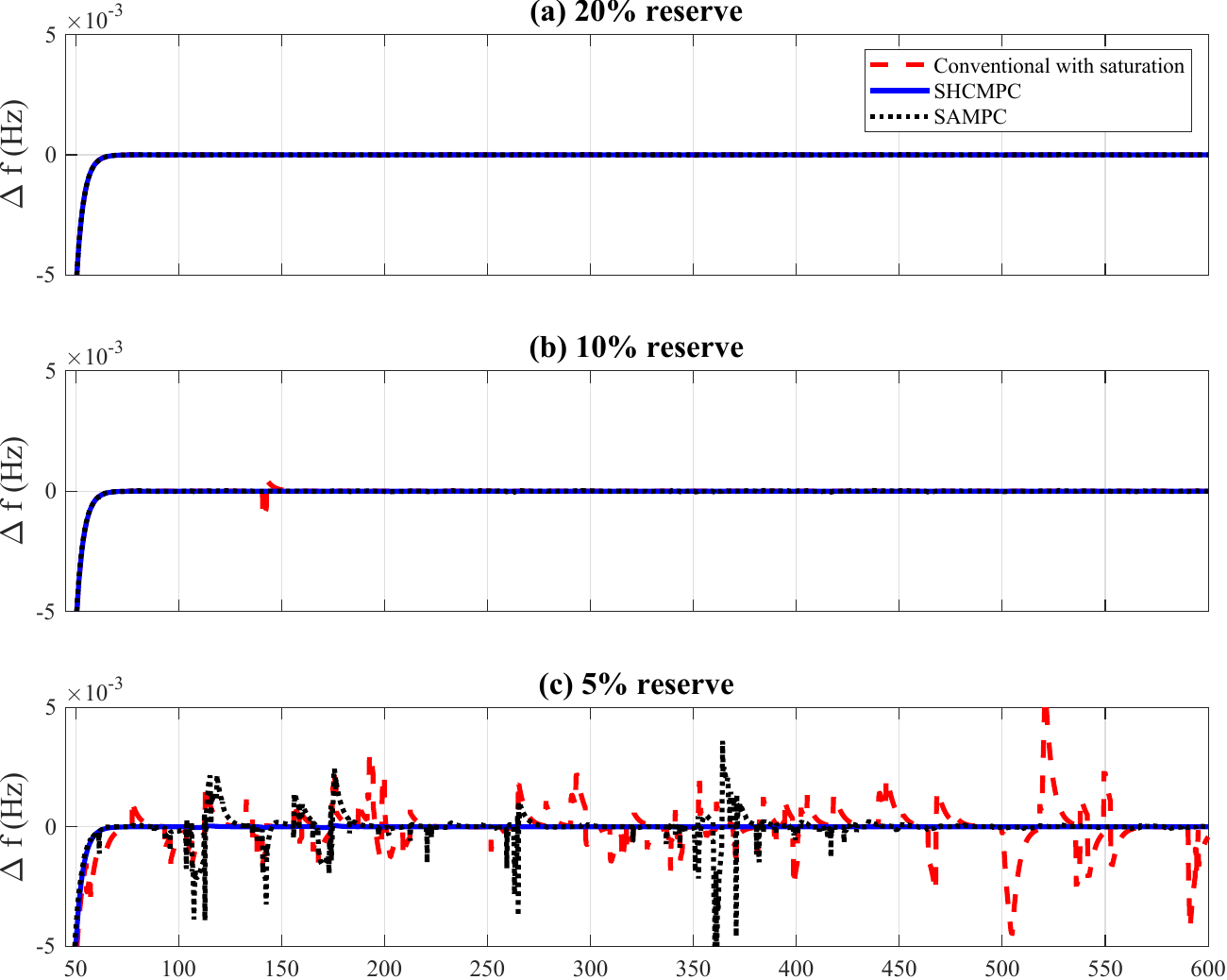}
    \caption{Frequency response of conventional MPC with stochastic saturation, SHCMPC, and SAMPC under different reserve margins.}
    \label{fig:headroom_aware_response}
    \vspace{-0.5cm}
\end{figure}

This subsection evaluates SHCMPC and SAMPC under the same reserve-margin
scenarios, load disturbance, headroom prediction, and physical saturation model
used in the previous subsection. Fig.~\ref{fig:headroom_aware_response}
focuses on the post-transient frequency deviations, where stochastic headroom
variations continue to affect the delivered regulation power. At the \(20\%\)
reserve margin, all responses are nearly identical, while small deviations
appear in the conventional saturated response at \(10\%\). The difference becomes
significant in the \(5\%\) reserve case, where conventional MPC exhibits
persistent fluctuations, whereas both headroom-aware strategies substantially
reduce them.

The improvement is due to the different ways the two methods use headroom
information. SHCMPC enforces predicted headroom through time-varying input
bounds, avoiding infeasible commands and reallocating regulation effort among
available PV units. SAMPC instead uses adaptive input penalties to discourage
reliance on PV units with limited predicted headroom, reducing stochastic
deviations without imposing hard headroom constraints. The following subsection
quantitatively compares these strategies in terms of LFC performance,
saturation mitigation, and online computation time.

\subsection{Comparative Performance Evaluation}

The three control schemes are evaluated from three complementary perspectives. First, LFC performance is assessed using the integral frequency error indices defined previously, which quantify how effectively each LFC scheme regulates the frequency deviation. Second, saturation mitigation is evaluated by examining the extent and duration of stochastic saturation under each control strategy. This comparison indicates how effectively each scheme accounts for limited and time-varying PV headroom during control allocation. Third, online computational efficiency is considered because MPC-based LFC must be solved repeatedly in real time, and the computational burden becomes increasingly important for larger scale systems.

\begin{table}[t]
\centering
\caption{LFC performance comparison of stochastic-saturation MPC, SHCMPC, and SAMPC.}
\label{tab:lfc_performance_comparison}
\setlength{\tabcolsep}{3.5pt}
\begin{tabular}{c l c c c c}
\toprule
\textbf{Reserve} & \textbf{Case} & \textbf{IAE} & \textbf{ISE} & \textbf{ITAE} & \textbf{ITSE} \\
\midrule
\multirow{3}{*}{20\%}
& Stoch. Sat. & 0.3711 & 0.0201 & 16.14 & 0.8384 \\
& SHCMPC      & 0.3711 & 0.0201 & 16.14 & 0.8382 \\
& SAMPC       & 0.3728 & 0.0203 & 16.65 & 0.8460 \\
\midrule
\multirow{3}{*}{10\%}
& Stoch. Sat. & 0.3882 & 0.0231 & 17.19 & 0.9591 \\
& SHCMPC      & 0.3849 & 0.0230 & 16.71 & 0.9587 \\
& SAMPC       & 0.3759 & 0.0218 & 17.62 & 0.9042 \\
\midrule
\multirow{3}{*}{5\%}
& Stoch. Sat. & 0.7100 & 0.0425 & 122.19 & 1.9009 \\
& SHCMPC      & 0.4084 & 0.0416 & 17.65  & 1.7011 \\
& SAMPC       & 0.4920 & 0.0393 & 44.75  & 1.6416 \\
\bottomrule
\end{tabular}
\end{table}

Table~\ref{tab:lfc_performance_comparison} shows that the benefit of headroom-aware control becomes more pronounced as the scheduled reserve margin decreases. At the \(20\%\) reserve margin, all three methods yield similar LFC performance, indicating that stochastic saturation has limited impact when sufficient headroom is available. The slightly larger error indices of SAMPC in this case are mainly due to its unconstrained adaptive penalty based formulation and the resulting input weighting differences, rather than saturation effects. At the \(10\%\) reserve margin, the headroom-aware schemes begin to improve the accumulated frequency error metrics. The most significant improvement occurs in the \(5\%\) reserve case, where SHCMPC reduces the IAE and ITAE by approximately \(42\%\) and \(86\%\), respectively, relative to the stochastic-saturation baseline. SAMPC also provides substantial improvement, reducing the IAE and ITAE by approximately \(31\%\) and \(63\%\), respectively. Although SHCMPC generally provides stronger suppression of persistent frequency deviations, SAMPC achieves comparable performance in several squared-error metrics, indicating that its adaptive allocation reduces the magnitude of large deviations even when some residual errors persist. Overall, these results show that headroom-aware LFC can maintain acceptable regulation performance under tighter reserve margins, suggesting a potential reduction in scheduled reserve requirements and improved economic utilization of PV resources.

\begin{table}[t]
\centering
\caption{Saturation mitigation and computational efficiency comparison.}
\label{tab:saturation_computation_comparison}
\setlength{\tabcolsep}{3.0pt}
\renewcommand{\arraystretch}{0.95}
\resizebox{\columnwidth}{!}{%
\begin{tabular}{c l c c c}
\toprule
\textbf{Reserve} & \textbf{Case} 
& \makecell{\textbf{Total Sat.}\\\textbf{Severity}} 
& \makecell{\textbf{Sat. Duration}\\\textbf{(s)}} 
& \makecell{\textbf{Avg. Sol.}\\\textbf{Time (ms)}} \\
\midrule
\multirow{3}{*}{20\%}
& Stoch. Sat. & 0.45 & 0.1   & 0.616 \\
& SHCMPC      & 0    & 0     & 0.623 \\
& SAMPC       & 1.28 & 0.1   & 0.202 \\
\midrule
\multirow{3}{*}{10\%}
& Stoch. Sat. & 2.07 & 2.3   & 0.671 \\
& SHCMPC      & 0    & 0     & 0.696 \\
& SAMPC       & 6.87 & 0.4   & 0.212 \\
\midrule
\multirow{3}{*}{5\%}
& Stoch. Sat. & 1667.47 & 526.3 & 0.680 \\
& SHCMPC      & 0       & 0     & 0.683 \\
& SAMPC       & 171.14  & 150.4 & 0.208 \\
\bottomrule
\end{tabular}%
}
\vspace{-0.6cm}
\end{table}

Table~\ref{tab:saturation_computation_comparison} reports the total saturation severity and saturation duration computed from \eqref{eq:sat_severity} and \eqref{eq:sat_duration}, respectively. The results show that SHCMPC eliminates saturation in all reserve margin cases, which explains its stronger LFC performance under tight reserve conditions. SAMPC does not remove saturation completely; however, its saturation behavior is substantially moderated in the most constrained case. For the $5\%$ reserve margin, SAMPC reduces the total saturation severity from $1667.47$ to $171.14$, corresponding to an approximately $90\%$ reduction, and decreases the saturation duration from $526.3s$  to $150.4s$ , corresponding to an approximately $71\%$ reduction. These results indicate that SHCMPC provides the strongest saturation mitigation, while SAMPC significantly limits the magnitude and persistence of saturation relative to the conventional stochastic-saturation baseline.

Although SHCMPC provides the strongest LFC performance and fully eliminates saturation, this improvement is obtained with a higher online computational burden due to the time-varying hard headroom constraints. In contrast, SAMPC incorporates headroom awareness through adaptive input weighting, which leads to a computationally lighter optimization problem. As shown in Table~\ref{tab:saturation_computation_comparison}, SAMPC requires an average solution time of approximately \(0.20\)-\(0.21~\mathrm{ms}\) across different reserve margin scenarios, whereas SHCMPC requires approximately \(0.62\)-\(0.70~\mathrm{ms}\). Thus, SAMPC achieves roughly a threefold reduction in average solution time while still providing substantial saturation mitigation and competitive LFC performance, especially under tight reserve conditions.

\section{Conclusion}
\label{Sec_7}

This paper investigated the impact of short-term renewable intermittency on
MPC-based LFC in IBR-dominated microgrids. The results show that fixed-headroom
MPC can overestimate the available regulation capability when renewable
generation capacity varies over short time scales. In this case, stochastic PV
headroom variations may cause optimized LFC commands to exceed physically
deliverable power, producing an optimization-actuation mismatch referred to as
stochastic saturation. This effect becomes more pronounced under reduced reserve
margins, showing that short-term headroom variability should be considered when
IBRs serve as primary LFC resources.

To mitigate this issue, two headroom-aware MPC strategies were developed.
SHCMPC incorporates predicted headroom through time-varying input constraints
and eliminates saturation in the tested scenarios, providing the strongest
suppression of persistent frequency deviations. SAMPC instead embeds headroom
awareness through adaptive input penalties, achieving substantial saturation
mitigation and LFC improvement with significantly lower computation time.
Overall, the comparative results highlight a tradeoff between strict saturation
avoidance and online computational efficiency, and show that incorporating
short-term headroom information into MPC-based LFC can improve renewable
utilization under reduced reserve margins while maintaining reliable frequency
regulation.

\vspace{-0.5cm}

\bibliographystyle{ieeetr}
\bibliography{references}
\end{document}